\documentclass[twocolumn,prb,superscriptaddress,showpacs,epsf]{revtex4}

\usepackage{graphicx}

\begin{document}
\title{Fast two-bit operations in inductively coupled flux qubits}
\date{\today}
\author{J. Q. You}
\altaffiliation[Email:~jqyou@fudan.edu.cn]{}
\affiliation{Frontier Research System, The Institute of Physical
and Chemical Research (RIKEN), Wako-shi 351-0198, Japan}
\affiliation{Department of Physics and National Laboratory for Surface Physics,
Fudan University, Shanghai 200433, China}
\author{Y. Nakamura}
\altaffiliation[Email:~yasunobu@frl.cl.nec.co.jp]{}
\affiliation{Frontier Research System, The Institute of Physical
and Chemical Research (RIKEN), Wako-shi 351-0198, Japan}
\affiliation{NEC Fundamental and Environmental Research
Laboratories, Tsukuba, Ibaraki 305-8051, Japan}
\affiliation{CREST, Japan Science and Technology Agency (JST),
Kawaguchi, Saitama 332-0012, Japan}
%
\author{Franco Nori}
\altaffiliation[Email:~nori@umich.edu]{}
\affiliation{Frontier Research System, The Institute of Physical
and Chemical Research (RIKEN), Wako-shi 351-0198, Japan}
\affiliation{Center for Theoretical Physics, Physics Department,
Center for the Study of Complex Systems,
University of
Michigan, Ann Arbor, MI 48109-1120, USA}

\begin{abstract}
A central problem for implementing efficient quantum computing is how to realize
fast operations (both one- and two-bit ones).
However, this is difficult to achieve for a collection of qubits, especially for those
separated far away,
because the interbit coupling is usually much weaker than the intrabit coupling.
Here we present an experimentally feasible method to effectively couple two flux qubits
via a common inductance and treat both single and coupled flux qubits with more realistic
models which include the loop inductance.
The main advantage of our proposal is that
a strong interbit coupling can be achieved using a small inductance,
so that two-bit operations as fast as one-bit ones can be easily realized.
We also show the flux dependence of the transitions between states for the coupled
flux qubits.

\end{abstract}
\maketitle

\section{Introduction}

Josephson-junction circuits can exhibit quantum behaviors.
Among qubits based on Josephson-junction circuits, the charge qubit realized in
a Cooper-pair box can demonstrate quantum oscillations.\cite{NAKA}
An improved version of this circuit has showed quantum oscillations
with a high quality factor.\cite{VION}
In addition to charge qubits,
flux qubits achieved in a superconducting
loop with one~\cite{FRIED} or three Josephson junctions~\cite{MOOIJ} have been studied
and some of these have shown quantum dynamics.\cite{CHIO} The phase qubit
consists of a large-area current-biased Josephson junction.\cite{HAN}

Capacitive couplings of two superconducting qubits (both charge-~\cite{PASH} and
phase-types~\cite{BERK}) were attained recently in experiments,
and quantum entanglement was observed in these systems.
Also, controllable interbit couplings of charge qubits
were proposed using a variable electrostatic transformer,\cite{AVERIN}
a current-biased Josephson junction~\cite{BLAIS} and a tunable dc-SQUID.\cite{SIE}
These interbit couplings can link nearest neighboring qubits.
Actually, there are quantum-computing protocols
(e.g., adiabatic quantum computing\cite{KAM}) that only demand nearest-neighbor couplings.
However, for more general quantum-computing protocols,
it is desirable to achieve strong enough couplings among non-neighboring qubits as well.
When charge qubits are coupled by $LC$-oscillator modes~\cite{MAK}
or by an inductance,\cite{YOU} long-ranged interbit couplings can be realized,
but a very large value of the inductance is needed.
An alternative way of coupling charge qubits was proposed using a Josephson
junction.\cite{YTN,LAN,SUN}
Moreover, the charge qubit can be very sensitive to the background charge fluctuations,
which generate noise that severely limits the performance of charge-qubit devices
and, unfortunately, is difficult to reduce.

In this paper, we present an experimentally feasible method to
effectively couple two flux qubits.
In contrast with the charge qubit, the flux qubit is {\it insensitive} to the charge noise.
In this qubit, the major noise is due to the fluctuations of the magnetic fluxes.
Estimations show that the flux qubit can have a relatively high quality
factor.\cite{WAL} Here we include the effect of the loop inductance in a three-junction
flux qubit and couple two flux qubits via a common inductance.
Because the critical current of each Josephson junction in the flux qubit is larger than
that in the charge qubit, we can produce a {\it strong} interbit coupling
using an inductance as small as 20~pH (corresponding to a loop diameter of 
approximately $16~\mu$m and comparable to the loop inductance of
the single flux qubit currently achieved in experiments),
and thereby two-bit operations as fast as one-bit ones can be easily achieved,
improving the efficiency of quantum computing.
Moreover, we show a novel flux dependence of the state transitions
in two coupled flux qubits.
We find that, except for some specific values of the external flux,
the forbidden transitions in the two coupled flux qubits become allowed
when the parameters of the two qubits change from being initially equal to each other 
and then making these different.

Coupling two flux qubits by a mutual inductance was proposed in Refs.~\onlinecite{ORL,SW,STO}
and was recently realized in experiments.\cite{MAJ,IZM}
Here we treat both single and coupled flux qubits using more realistic models
which include the loop inductance.
We numerically solve the Schr{\"o}dinger equation to obtain
the energy levels and the eigenstates of the flux-qubit systems.
This numerical method allows us to extend our study to the larger inductance regime.

The paper is organized as follows. In Sec.~II, we study a single flux qubit
containing loop inductance. It is shown that the system
can still be used to achieve a qubit even for a larger loop inductance of $L\sim 1$~nH.
Section III focuses on two flux qubits coupled by a common inductance.
In Sec.~IV, we study the state transitions induced by the microwave field.
Section V deals with the circulating supercurrents and quantum measurement.
Finally, the discussion and conclusion are given in Sec.~VI.

\begin{figure}
\includegraphics[width=3.4in,
bbllx=37,bblly=468,bburx=570,bbury=702]{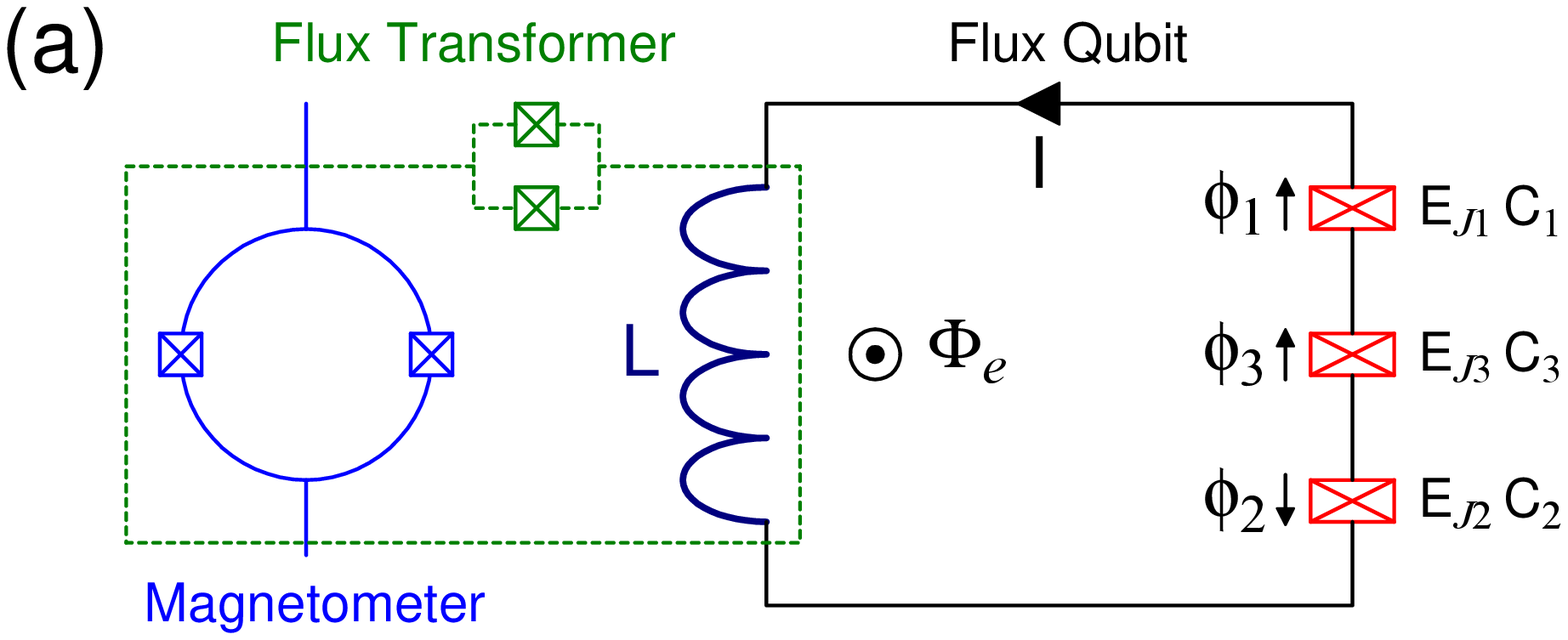}
\includegraphics[width=3.4in,
bbllx=36,bblly=160,bburx=571,bbury=473]{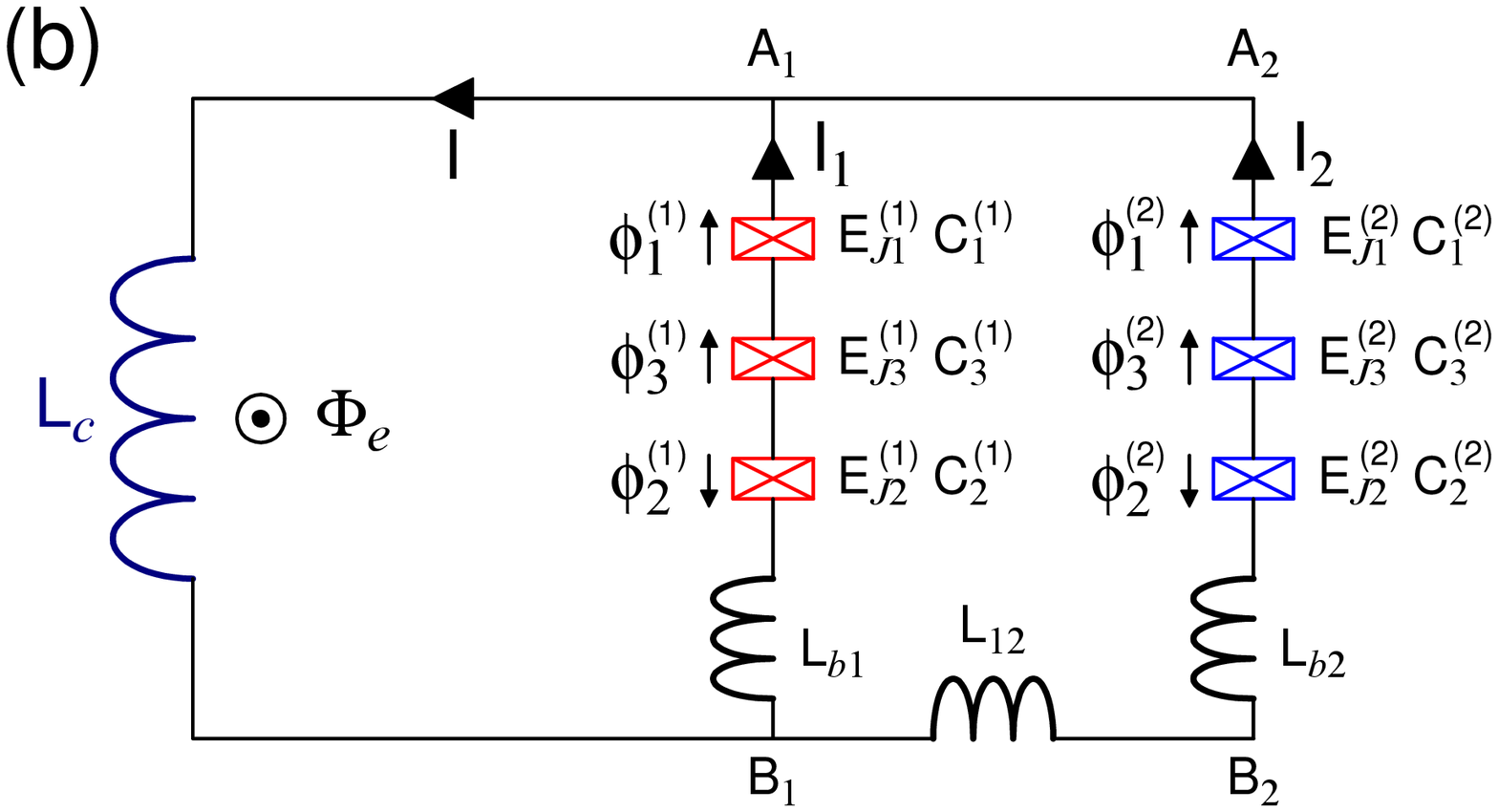}
\caption{(Color online) (a)~A flux qubit, where an external magnetic flux $\Phi_e$ pierces the
superconducting loop that contains three Josephson junctions and an inductance $L$.
The Josephson energies and capacitances of the junctions are $E_{J1}=E_{J2}=E_J$,
$C_1=C_2=C$, $E_{J3}=\alpha E_J$, and $C_3=\alpha C$. Here we choose $\alpha=0.8$
and $E_J=35E_c$, where $E_c=e^2/2C$.
(b)~Two flux qubits coupled by a common inductance $L_c$, where the external flux $\Phi_e$
is applied within the left loop $A_1L_cB_1A_1$. The parameters of each flux qubit are
$E_{J1}^{(i)}=E_{J2}^{(i)}=E_J^{(i)}$, $C_1^{(i)}=C_2^{(i)}=C^{(i)}$,
$E_{J3}^{(i)}=\alpha_i E_J^{(i)}$, and $C_3^{(i)}=\alpha_i C^{(i)}$, with $i=1,2$.
Here we choose $\alpha_i=0.8$ and $E_J^{(i)}=35E_c^{(i)}$,
where $E_c^{(i)}=e^2/2C^{(i)}$.
To implement a readout of the flux-qubit states, a switchable superconducting flux
transformer is employed to couple the dc-SQUID magnetometer with the inductance $L$ in (a)
or $L_c$ in (b) during the quantum measurement. However, this coupling is switched off
in the absence of a readout.
}
\label{fig1}
\end{figure}

\section{Single flux qubit}

\subsection{The model}

We first consider a single flux qubit in the absence of a quantum measurement,
where the dc-SQUID magnetometer for measuring quantum states of the flux qubit
is decoupled from the qubit.
As shown in Fig.~1(a), the flux qubit consists of a superconducting loop with three
Josephson junctions and the total inductance of the whole loop is $L$.
Fluxoid quantization around the loop imposes a constraint on the phase drops across
the three junctions:
\begin{equation}
\phi_1-\phi_2+\phi_3+2\pi f'=0,
\end{equation}
where
\begin{equation}
f'=f+{IL\over\Phi_0}.
\end{equation}
Here, $\Phi_0=h/2e$ is the flux quantum, 
\begin{equation}
f={\Phi_e/\Phi_0} 
\end{equation}
represents the reduced magnetic
flux, and
\begin{equation}
I=I_0\sin\phi_1,
\end{equation}
with
$I_0=2\pi E_J/\Phi_0$,
is the circulating supercurrent.

When the loop inductance is included, the Hamiltonian of the single flux qubit is
\begin{equation}
H={P_p^2\over 2M_p}+{P_m^2\over 2M_m}+U(\phi_p,\phi_m),
\end{equation}
with the potential energy given by
\begin{eqnarray}
U(\phi_p,\phi_m)\!&\!=\!&\!E_J[2+\alpha-2\cos\phi_p\cos\phi_m \nonumber\\
&&-\alpha\cos(2\pi f'+2\phi_m)]+{1\over 2}LI^2.
\end{eqnarray}
Here 
\begin{eqnarray}
&&P_k=-i\hbar{\partial\over\partial\phi_k},\;\;\;k=p, m,\nonumber\\
&&M_p=2C(\Phi_0/ 2\pi)^2,\\
&&M_m=M_p(1+2\alpha),\nonumber
\end{eqnarray} 
and 
\begin{eqnarray}
&&\phi_p={1\over 2}(\phi_1+\phi_2),\nonumber\\
&&\phi_m={1\over 2}(\phi_1-\phi_2). 
\end{eqnarray}
Also, the supercurrent $I$ can be rewritten as
\begin{equation}
I=I_0\sin(\phi_p+\phi_m).
\end{equation}
The Hamiltonian (5) is reduced to Eq.~(12) in Ref.~\onlinecite{ORL} when $L\rightarrow 0$.

\begin{figure}
\includegraphics[width=3.4in,
bbllx=101,bblly=338,bburx=486,bbury=720]{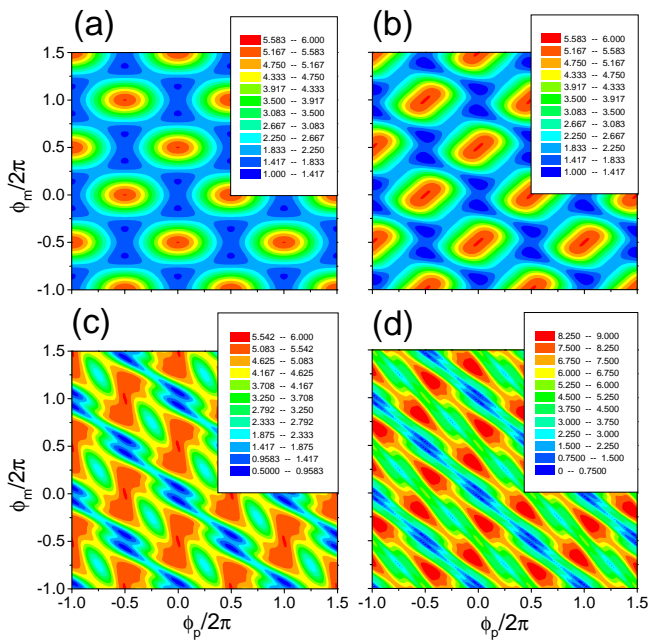}
\caption{(Color online) Contour plots of the potential energy $U(\phi_p,\phi_m)$, in units of $E_J$,
for $\alpha=0.8$
and $f=0.5$. Here $\beta_L\equiv 2\pi I_0L/\Phi_0=$ (a) 0, (b) 1, (c) 4, and (d) 10.
Notice that the well defined double-well potential structure vanishes in (d),
and thus the flux qubit breaks down.}
\label{fig2}
\end{figure}

Figure 2 presents the contour plots of the periodic potential $U(\phi_p,\phi_m)$
for $f=0.5$ and $\alpha=0.8$.
The numerical results show that the minima of the potential {\it preserve}
the two-dimensional centered cubic lattice even for a large loop inductance.
For inductance ratio 
\begin{equation}
\beta_L\equiv L/L_J
\end{equation}
from zero to one (where 
$L_J=\Phi_0/2\pi I_0$ is the Josephson-junction inductance),
a well defined double-well potential structure exists at each lattice point
even though at higher energies the well shapes are modified by
the loop inductance $L$.
This double-well structure is required for achieving a two-level system.
As shown in Fig.~3, the lowest two levels of the single-qubit system
are not significantly affected by
the variation of $\beta_L$ (when $0\leq \beta_L\alt 1$)
because the corresponding two eigenstates are mainly
contributed by the weakly $\beta_L$-dependent ground state in each well.
However, since varying $L$ significantly modifies the well shapes at higher energies,
the excited states within or above the wells (which, as seen in Fig.~3,
dominantly contribute to the eigenstates corresponding to the third and higher levels)
become pronouncedly $\beta_L$-dependent.
Indeed, Figure~3 shows that the top three levels
are sensitive to the variation of $\beta_L$ (even when $0\alt\beta_L\alt 1$).
Moreover, with the loop inductance increasing to $\beta_L\approx 4$ [see Fig.~2(c)],
a more distorted double-well structure appears at each lattice point,
and a local energy minimum develops along the diagonal direction between
every two adjoining double-well structures.
These newly-developed local minima will affect the two-level system achieved for
the qubit. When the loop inductance increases even more
to $\beta_L\approx 10$ [see Fig.~2(d)], the periodic potential is even more distorted.
In this case, the well defined double-well potential structure vanishes,
and thus the flux qubit breaks down.

\begin{figure}
\includegraphics[width=3.3in,
bbllx=31,bblly=106,bburx=554,bbury=724]{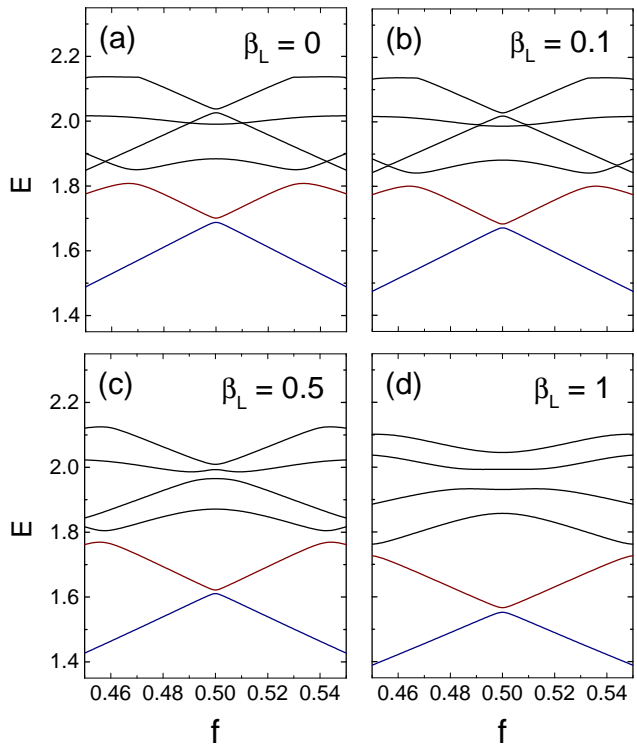}
\caption{(Color online) Energy levels of a single flux qubit versus reduced flux $f$
for different values of $\beta_L$, where only the levels of
the states $|i\rangle$, $i=0$ to 5, are shown.
Here the energy $E$ is in units of $E_J$.
Notice the robustness of the two lowest levels for wide changes in the
loop inductance $L$.}
\label{fig3}
\end{figure}

\subsection{Energy spectrum}

The energy spectrum and the eigenstates are
determined by
\begin{equation}
H\Psi(\phi_p,\phi_m)=E\Psi(\phi_p,\phi_m).
\end{equation}
Figure~3 shows the dependence of the energy levels on the magnetic flux
for $\beta_L\leq 1$. Here we choose $E_J=35E_c$, where
the charging energy $E_c$ is defined as $E_c=e^2/2C$. These parameters are close to
those used in a recently fabricated flux-qubit device.\cite{CHIO}

Around $f=0.5$, in sharp contrast with the higher energy levels, the energy difference
\begin{equation}
\Delta=\varepsilon_1-\varepsilon_0
\end{equation}
between the lowest two levels is {\it not} sensitive to the variation of $\beta_L$.
In Fig.~4, we show the energy separation of the two lowest levels, $\Delta$,
as a function of $\beta_L$.
We find the interesting result that $\Delta(\beta_L)$ is almost flat at $f=0.5$
($0.011<\Delta(\beta_L)/E_J<0.0135$) when $0\leq \beta_L \leq 0.85$.
These features indicate that, even with a large loop inductance of $\beta_L=1$,
in the vicinity of $f=0.5$ the two lowest eigenstates (denoted by $|0\rangle$
and $|1\rangle$ for the ground and the first excited states, respectively) remain 
suitable basis states for a flux qubit.
Within the subspace of qubit states spanned by $|0\rangle$ and $|1\rangle$,
the Hamiltonian is reduced to
\begin{equation}
H=\varepsilon_1|1\rangle\langle 1|+\varepsilon_0|0\rangle\langle 0|.
\end{equation}
If the average energy $(\varepsilon_1+\varepsilon_0)/2$ is chosen to be the
new zero-point energy of the flux qubit, the Hamiltonian can be further expressed as
\begin{equation}
H={1\over 2}\Delta\rho_z,
\end{equation}
where $\rho_z=|1\rangle\langle 1|-|0\rangle\langle 0|$.

\begin{figure}
\includegraphics[width=3.1in,
bbllx=59,bblly=225,bburx=541,bbury=571]{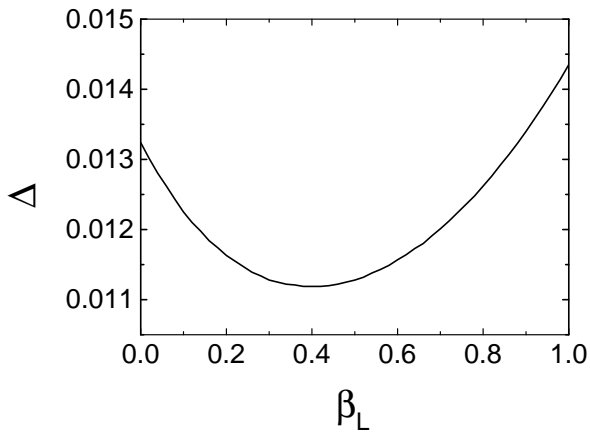}
\caption{Energy difference $\Delta$ between qubit states $|1\rangle$ and $|0\rangle$
as a function of $\beta_L$ for $f=0.5$. Here $\Delta$ is in units of $E_J$.
Notice that the energy difference varies $\sim 0.003E_J$ when varying 
the loop inductance $L$.}
\label{fig4}
\end{figure}

\subsection{Comparision with other works}

In Ref.~\onlinecite{CRA},
the effects of the loop inductance in a flux qubit are considered using
a perturbation approach, where the Hamiltonian is expanded into
three parts: an inductance-free Hamiltonian, an inductance-related harmonic oscillator term,
and a small correction term.
This perturbation method is valid for $\beta_L\ll 1$ since the correction term is
proportional to the loop inductance of the flux qubit.
Instead of using the perturbation approach,
we numerically solve Eq.~(11) to obtain the eigenvalues and eigenstates of the system.
This numerical method allows us to extend our study to
the regime of $\beta_L\sim 1$,
where the lowest two eigenstates of the system can still be used for achieving a qubit.
Using the experimental value\cite{CHIO} $I_0\sim 0.5$~$\mu$A,
this regime corresponds to a loop inductance of $L\sim 1$~nH.

\section{Coupled flux qubits}

\subsection{The model}

To couple two flux qubits, we use a {\it common}
inductance $L_c$ shared by these two qubits [see Fig.~1(b)].
Here the external flux $\Phi_e$ is applied within the loop $A_1L_cB_1A_1$.
Also, the circuit is designed in such a way that the mutual inductance
between loops $A_1L_cB_1A_1$ and $A_1B_1B_2A_2A_1$ may be ignored.
This is achieved when only a small fraction of the flux generated by one loop
passes through the other.
(If this were not to be the case, the interbit coupling can still be achieved by the
common inductance $L_c$, but the interaction Hamiltonian takes a more complicated form.)
Phase drops through the three Josephson junctions
of the $i$th flux qubit are constrained by
\begin{equation}
\phi_1^{(i)}-\phi_2^{(i)}+\phi_3^{(i)}
+2\pi \left[f+(I_iL_i+I_jL_c)/\Phi_0\right]=0,
\end{equation}
where $i,j=1,2$ ($i\ne j$), and 
\begin{eqnarray}
&&L_1=L_c+L_{b1},\nonumber\\ 
&&L_2=L_c+L_{12}+L_{b2}.
\end{eqnarray}
The total supercurrent through $L_c$ is
\begin{equation}
I=I_1+I_2,
\end{equation}
where
\begin{equation}
I_i=I_{0i}\sin(\phi_{pi}+\phi_{mi}),
\end{equation}
with 
$I_{0i}={2\pi E_J^{(i)}/\Phi_0}$,
and
\begin{eqnarray}
&&\phi_{pi}={1\over 2}(\phi_1^{(i)}+\phi_2^{(i)}),\nonumber\\
&&\phi_{mi}={1\over 2}(\phi_1^{(i)}-\phi_2^{(i)}).
\end{eqnarray}

The Hamiltonian of the two coupled flux qubits can be written as
\begin{equation}
H=H_1+H_2+H_I.
\end{equation}
Here $H_i$ is the Hamiltonian of the $i$th isolated flux qubit, with loop inductance
$L_i$ and circulating supercurrent $I_i$, which has the form in Eq.~(5) but with
$f'$ replaced by
\begin{equation}
f'_i=f+{I_iL_i\over\Phi_0}.
\end{equation}
Also, $E_J$, $C$ and $\alpha$  are replaced by $E_J^{(i)}$,
$C^{(i)}$ and $\alpha_i$.
The interaction Hamiltonian is
\begin{equation}
H_I=L_cI_1I_2-\sum_{i=1}^2\alpha_iE_J^{(i)}\Pi_i,
\end{equation}
where
\begin{equation}
\Pi_i=\cos(\gamma_i+2\pi I_jL_c/\Phi_0)-\cos\gamma_i,
\end{equation}
with
\begin{equation}
\gamma_i=2\pi f'_i+2\phi_{mi},
\end{equation}
and $i\ne j$.

When $\beta_{Li}\equiv 2\pi I_{0i}L_c/\Phi_0\ll 1$, $H_I$ is approximated by
\begin{equation}
H_I=-L_cI_1I_2
\end{equation}
because 
\begin{equation}
\alpha_iE_J^{(i)}\Pi_i\approx L_cI_1I_2 
\end{equation}
in this case (see the Appendix).
Within the qubit-state subspace of the $i$th isolated flux qubit, $H_i$ is reduced to
\begin{equation}
H_i={1\over 2}\Delta_i\rho_z^{(i)},
\end{equation}
where 
\begin{equation}
\rho_z^{(i)}=|1_i\rangle\langle 1_i|-|0_i\rangle\langle 0_i|.
\end{equation}
In the vicinity of $f=0.5$, because the supercurrents $I_i$ at states $|1_i\rangle$
and $|0_i\rangle$ have equal magnitudes but opposite directions,
$I_i$ can be written as
\begin{equation}
I_i=a_i\rho_z^{(i)}+b_i|1_i\rangle\langle 0_i|+b^*_i|0_i\rangle\langle 1_i|,
\end{equation}
where 
\begin{eqnarray}
&& a_i=\langle 1_i|I_i|1_i\rangle,\nonumber\\
&& b_i=\langle 1_i|I_i|0_i\rangle.
\end{eqnarray}
Because the supercurrent $I_i$ at state $|1_i\rangle$ (i.e., $a_i$) is proportional to
the slope of the energy level that corresponds to state $|1_i\rangle$ with respect to $f$
(see, e.g., Ref.~\onlinecite{ORL}), it falls to zero at the {\it symmetric} point, $f=0.5$,
where the level becomes flat.
Also, our numerical results show that $b_i$ becomes a real number
at $f=0.5$. Thus, we can rewrite $I_i$ at $f=0.5$ as
\begin{equation}
I_i=b_i\rho_x^{(i)},
\end{equation}
with 
\begin{equation}
\rho_x^{(i)}=|1_i\rangle\langle 0_i|+|0_i\rangle\langle 1_i|.
\end{equation}
For $\beta_{Li}\ll 1$, i.e., the common inductance $L_c$ is very small, 
the Hamiltonian at $f=0.5$ can be cast to
\begin{equation}
H=\sum_{i=1}^2{1\over 2}\Delta_i\rho_z^{(i)}-\chi\rho_x^{(1)}\rho_x^{(2)},
\end{equation}
with 
\begin{equation}
\chi=L_cb_1b_2.
\end{equation}
It is clear that the interbit coupling persists at $f=0.5$.

\begin{figure}
\includegraphics[width=3.4in,
bbllx=55,bblly=129,bburx=549,bbury=663]{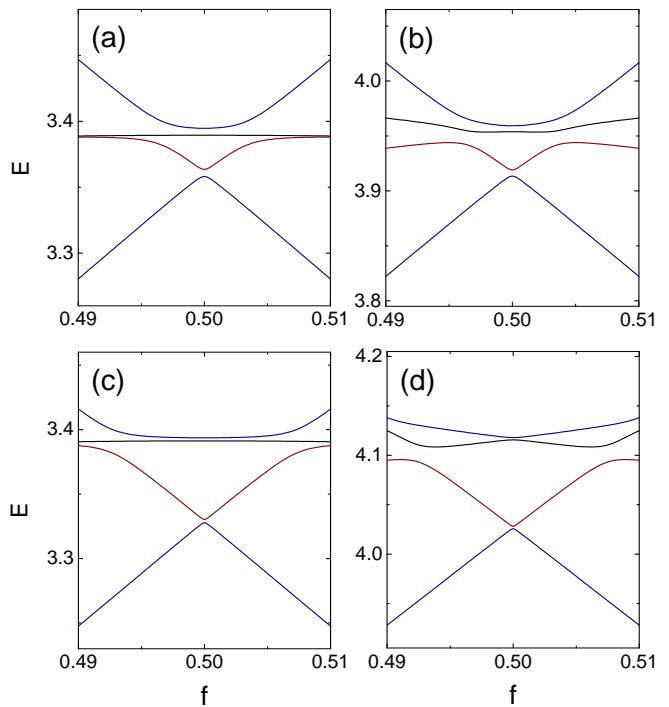}
\caption{(Color online) Energy levels of two coupled flux qubits versus reduced flux $f$ for
$L_{b1}/L_c=0.1$ and $(L_{12}+L_{b2})/L_c=0.2$.
The parameters $\beta_{Li}\equiv 2\pi I_{0i}L_c/\Phi_0$ are (a) $\beta_{L1}=\beta_{L2}=0.03$,
(b) $\beta_{L1}=0.03$, $\beta_{L2}=0.04$, (c) $\beta_{L1}=\beta_{L2}=0.07$, and
(d) $\beta_{L1}=0.07$, $\beta_{L2}=0.1$. Here the energy $E$ is in units of $E_J^{(1)}$.
Near $f=0.5$, the energy levels are robust with respect to the asymmetry between 
$\beta_{L1}$ and $\beta_{L2}$.}
\label{fig5}
\end{figure}

\subsection{Energy spectrum}

Figure~5 shows the energy spectrum of the two coupled flux qubits around
$f=0.5$. In order to realize fast two-bit operations while keeping
the leakage from the qubit states to other higher energy states small, we choose
the interbit coupling strength to be comparable to the energy difference, at $f=0.5$,
between the basis states $|1_i\rangle$ and $|0_i\rangle$ of each qubit.
As shown in Figs. 5(a) and 5(b),
the energy spectrum remains similar in the vicinity of $f=0.5$ when the two flux qubits have
different values of parameters.
Furthermore, the two higher energy levels, $\epsilon_3$ and $\epsilon_4$,
in the first four energy levels (i.e., $\epsilon_k$ with $k=1$ to 4)
of the two coupled flux qubits
are flat in a relatively broad range around $f=0.5$; this flat region is
{\it much broader} than the corresponding flat-energy-level range of the
single flux qubit around $f=0.5$.
The flux-independent level $\epsilon_3$ in Fig.~5(a) corresponds to a singlet eigenstate,
while other three levels correspond to triplet eigenstates. As expected, the transitions
between this singlet state and other three triplet states are not allowed
by the microwave perturbation [cf. Fig.~6(a)].

When the interbit coupling increases further, the flat region for both levels
$\epsilon_3$ and $\epsilon_4$ widens for two qubits
having identical parameters [see Fig.~5(c)],
but $\epsilon_3$ and $\epsilon_4$ become much different in this region
when the two qubits are not identical [see Fig.~5(d)].
Moreover, it can be seen that, at $f=0.5$, the gap between levels $\epsilon_1$ and 
$\epsilon_2$ and that between $\epsilon_3$ and $\epsilon_4$  become narrow
when increasing the interbit coupling.

At $f=0.5$, the first four energy levels, $\epsilon_k$, $k=1$ to 4,
of the coupled flux qubits can be approximated by
\begin{eqnarray}
&&\epsilon_1=-{1\over 2}E_A,\;\;\; \epsilon_3={1\over 2}E_B,\nonumber\\
&&\epsilon_2=-{1\over 2}E_B,\;\;\; \epsilon_4={1\over 2}E_A,
\end{eqnarray}
where
\begin{eqnarray}
&&E_A=[(\Delta_1+\Delta_2)^2+4\chi^2]^{1/2},\nonumber\\
&&E_B=[(\Delta_1-\Delta_2)^2+4\chi^2]^{1/2}.
\end{eqnarray}
The gap between levels $\epsilon_2$ and $\epsilon_3$ is $E_B$, which increases with $\chi$.
The gap between levels $\epsilon_1$ and $\epsilon_2$ and that between $\epsilon_3$ and
$\epsilon_4$ are given by $E_A-E_B$.
Figures 5(a) and 5(b) correspond to $\chi\approx\Delta_1$;
in Fig.~5(a) where $\Delta_1=\Delta_2=\Delta$,
the two equal gaps, $E_A-E_B$, at $f=0.5$ are $(\sqrt{2}-1)\Delta$.
When $\chi$ further increases, the value of $E_A-E_B$ decreases; namely, the two equal gaps
become narrow [cf. Figs.~5(c) and 5(d)].

In the case of Fig.~5(a), because $2\pi I_{0i}L_c/\Phi_0=0.03$, the common inductance is
\[
L_c\approx 20~{\rm pH} 
\]
if the critical currents $I_{0i}$ are equal to the experimental
value~\cite{CHIO} $I_0\sim 0.5$ $\mu$A. Such a small inductance is experimentally realizable,
e.g., using a loop of diameter $d\approx 16$~$\mu$m.
Also, our numerical calculations show that $b_i\approx 0.66I_0$ at $f=0.5$.
The interbit coupling is thus of the order 
\begin{equation}
\chi=L_cb_1b_2\approx 0.013E_J,
\end{equation}
which is equal to the energy difference $\Delta$ at $f=0.5$ of the single flux qubit
with $\beta_L=0.03$. 
The typical two-bit operation time related with the interbit-coupling strength 
is 
\begin{equation}
\tau_2\sim \hbar/\chi.
\end{equation}
Because $\chi\approx\Delta$ when $L_c\approx 20$~pH, 
\begin{equation}
\tau_2\sim \tau_1\equiv \hbar/\Delta,
\end{equation}
where $\tau_1$ is the one-bit operation time. Therefore,
the corresponding two-bit operation is as fast as the one-bit operation.

\subsection{Comparison with other works}

Spectral results similar to the ones shown in Figs.~5(a) and 5(b) were also obtained 
by Storcz and Wilhelm~\cite{STO} and by Majer {\it et al.}~\cite{MAJ} 
using simpler model Hamiltonians for two coupled flux qubits.
However, because a different setup is used in Ref.~\onlinecite{MAJ} for coupling 
the two qubits, the four energy levels are flipped as compared to ours.

In our proposed setup, the shared part of the loop is on the same side of each qubit
and, as given in Eq.~(37), $\chi >0$.
Since the negative coupling term in Eq.~(33) favors the parallel arrangement of 
the (pseudo)spins, the interbit coupling here is {\it ferromagnetic}. 
In contrast, in the setup of Ref.~\onlinecite{MAJ}, 
the shared loop is on the opposite side of each qubit.
This gives rise to a negative coupling parameter $\chi$. Therefore, the interbit coupling
there is {\it anti}ferromagnetic because the (pseudo)spins tend to 
arrange antiparallel to each other.

Here we use the {\it common} loop inductance to couple two flux qubits,
while different setups are proposed in Refs.~\onlinecite{ORL} and \onlinecite{SW}, 
where the {\it mutual} inductance is used for coupling flux qubits.
Since the mutual inductance is always smaller (sometimes can be much smaller) 
than the loop inductance of each flux qubit, if the two flux qubits are coupled 
via the mutual inductance (instead of the common loop inductance in our approach), 
a larger (sometimes much larger) loop 
is required for each qubit to produce a strong enough interbit coupling.
Therefore, due to the larger loop in each flux qubit, 
the system would experience more serious flux noise. 
This is a very significant difference between these approaches.

In our approach, we treat both single and coupled flux qubits using more realistic 
models including the loop inductance. 
If our more general theory were used to deal with the different setup proposed 
in, e.g., Ref.~\onlinecite{MAJ}, the simple model Hamiltonian used in 
Ref.~\onlinecite{MAJ} would be an approximation of the more general theory. 
The simpler model can explain the current experiment in Ref.~\onlinecite{MAJ},
and so does the general theory. Moreover, the more general theory presented here can  
explain additional features that would be relevant for future experiments, 
while the simple model might not. 
For instance, even for the single flux qubit, the simple model cannot 
tell how large the loop should be to break down the qubit. 
Moreover, the simple model involves only two levels for each qubit.
It cannot explain phenomena related with the state transitions 
from these two levels to higher ones. However, the general theory 
can do it.

\section{State transitions}

\subsection{Transition matrix elements for single and coupled two qubits}

When a microwave field with an appropriate
frequency $\omega$ is applied through the superconducting loop of the single flux qubit,
a transition between two states occurs. Now, the total flux within the loop is
$\Phi_e+\Phi_f$, where
\begin{equation}
\Phi_f(t)=\Phi_X\cos(\omega t+\theta)
\end{equation}
is the microwave-field-induced
flux through the loop.
For a weak microwave field, the single flux qubit experiences
a time-dependent perturbation
\begin{equation}
H'(t)=-I\Phi_X\cos(\omega t+\theta),
\end{equation}
and the transition matrix element $t_{ij}$ between states $|i\rangle$
and $|j\rangle$ is given by 
\begin{equation}
t_{ij}=\langle i|I\Phi_X|j\rangle.
\end{equation}
Similarly, when the microwave field is applied through the left loop $A_1L_cB_1A_1$
of the coupled flux qubits, the transition matrix element $t_{ij}$ between the
coupled-qubit states $|\epsilon_i\rangle$ and $|\epsilon_j\rangle$ is then
\begin{equation}
t_{ij}=\langle\epsilon_i|I\Phi_X|\epsilon_j\rangle.
\end{equation}
Notice the difference between Eqs. (42) and (43), though they look similar to each other
in expression; in Eq.~(42), $|i\rangle$ is an eigenstate of the single flux
qubit and $I$ is the circulating current in the qubit loop, 
while $I$ in Eq.~(43) is the total current $I=I_1+I_2$ in the coupled-qubit 
circuit and $|\epsilon_i\rangle$ is an eigenstate of the coupled two 
flux qubits.

\begin{figure}
\includegraphics[width=3.4in,
bbllx=50,bblly=31,bburx=549,bbury=669]{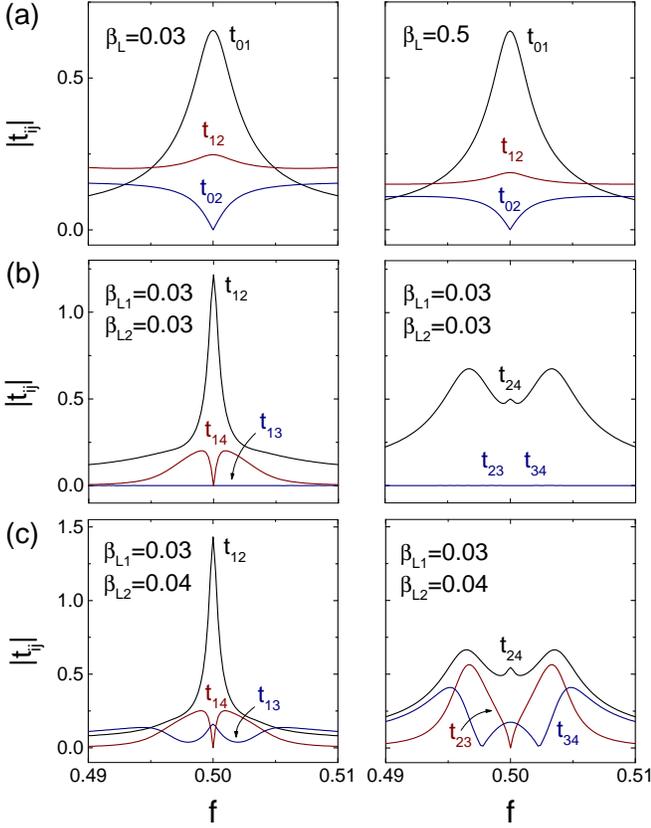}
\caption{(Color online) (a) Moduli of the transition matrix elements $t_{ij}$
between single-qubit states
$|i\rangle$ and $|j\rangle$ versus reduced flux $f$. (b) and (c) Moduli of
the transition matrix elements
between coupled-qubit states
$|\epsilon_i\rangle$ and $|\epsilon_j\rangle$ versus
$f$ for $L_{b1}/L_c=0.1$ and $(L_{12}+L_{b2})/L_c=0.2$.
Here $|t_{ij}|$ is in units of $I_{0}\Phi_X$ in (a) and $I_{01}\Phi_X$ in (b) and (c).
Notice that, in (b) and (c), some transitions for the coupled two qubits are 
sensitive with respect to the asymmetry of $\beta_{L1}$ and $\beta_{L2}$.}
\label{fig6}
\end{figure}

Figure~6(a) presents the flux dependence of $|t_{ij}|$ for transitions
$|0\rangle\rightarrow|1\rangle$, $|0\rangle\rightarrow|2\rangle$, and
$|1\rangle\rightarrow|2\rangle$ in a single flux qubit.
Because of the symmetry of the wave functions, $|t_{02}|=0$ at $f=0.5$, and
thus the transition $|0\rangle\rightarrow|2\rangle$ is forbidden.
Also, it can be seen that $|t_{01}|$ is not sensitive to the variation of $\beta_L$,
while $|t_{02}|$ and $|t_{12}|$ are slightly reduced when increasing $\beta_L$.
This observation is consistent with the energy spectrum in Fig.~3, where the gap between
the lowest two levels 0 and 1 is not significantly changed,
but the gap between levels 1 and 2 slightly increases with $\beta_L$.
In Figs.~6(b) and 6(c), we show the flux dependence of $|t_{ij}|$ for all possible
transitions in the coupled flux qubits.
When the two flux qubits have the same parameters, the transitions
\begin{eqnarray}
&&|\epsilon_1\rangle\rightarrow |\epsilon_3\rangle,\nonumber\\
&&|\epsilon_2\rangle\rightarrow |\epsilon_3\rangle,\\
&&|\epsilon_3\rangle\rightarrow |\epsilon_4\rangle \nonumber
\end{eqnarray}
are {\it forbidden}
because $|t_{ij}|=0$ [see Fig.~6(b)].
However, they are {\it allowed} (except for some specific values of $f$)
when the parameters of the two flux qubits are different [see Fig.~6(c)].
These properties are attributed to the changes in the symmetry of the qubit states . 
In particular, when $f=0.5$ , $|t_{12}|$ has the largest value, while $|t_{24}|$ has a
smaller value and others are either zero or much smaller.

\subsection{One- and two-bit operations implemented via microwave fields}

For a single flux qubit with $\beta_L=0.03$,
the energy difference $\Delta$ between states $|1\rangle$ and $|0\rangle$ is
$0.01291E_J$ at $f=0.5$.
Using an experimental value~\cite{CHIO} for the critical current $I_0\sim 0.5$~$\mu$A,
we obtain $E_J\sim 1.03$~meV.
The energy difference of $0.01291E_J$ corresponds to a gap of $\nu\approx 3.2$~GHz.
The one-bit operation can be implemented using a resonant microwave field.
For a weak driving field, the Rabi frequency $\Omega_{01}$ is given by $|t_{01}|/\hbar$.
The typical switching time is $t_{\rm SW}=\pi/\Omega_{01}$ when the states $|0\rangle$
and $|1\rangle$ {\it flip}. For instance, because $|t_{01}|\approx 0.66I_0\Phi_X$
at $f=0.5$, the switching time $t_{\rm SW}$ is about 3~ns for $I_0\Phi_X\sim 1$~$\mu$eV.
If the leakage from these two states to others is small, one can realize a fast
one-bit operation, e.g., with a switching time 
\[
t_{\rm SW}={\pi\over\Omega_{01}}\sim 10\nu^{-1}
(\approx 3~{\rm ns}), 
\]
by increasing the microwave-field intensity.

Let the energy difference between states $|0\rangle$ ($|1\rangle$) and $|1\rangle$
($|2\rangle$) be $\hbar\omega_{01}$ ($\hbar\omega_{12}$).
When the field is tuned to be {\it resonant} with the transition
$|0\rangle\rightarrow |1\rangle$, the ratio of the transition probabilities between
$|1\rangle\rightarrow |2\rangle$ and $|0\rangle\rightarrow |1\rangle$ can be estimated as
\begin{equation}
{\rho_{12}\over\rho_{01}}=\left(\Omega_{12}\over\Omega\right)^2
{\sin^2(\Omega\tau/2)\over\sin^2(\Omega_{01}\tau/2)},
\end{equation}
where
\begin{equation}
\Omega=[\Omega_{12}^2+(\omega_{12}-\omega_{01})^2]^{1/2},
\end{equation}
$\Omega_{12}=|t_{12}|/\hbar$, and $\tau$ is the duration of the microwave-field pulse.
When 
\[
\tau=\pi/\Omega_{01} \sim 10\nu^{-1},
\] 
using the numerical results
$\hbar\omega_{01}=0.01291E_J$, $\hbar\omega_{12}=0.18763E_J$, and
$|t_{12}/t_{01}|\approx 0.38$ at $f=0.5$, we have
\begin{equation}
{\rho_{12}\over\rho_{01}}\approx 1.5\times 10^{-6}.
\end{equation}
This implies that the leakage to other states is small
for a fast one-bit operation implemented via a microwave field.

Corresponding to Fig.~5(a), $|\epsilon_1\rangle$ and $|\epsilon_2\rangle$ at $f=0.5$
are approximated by
\begin{eqnarray}
|\epsilon_1\rangle\!&\!=\!&\!{1\over\sqrt{\eta^2+1}}(\eta\;|00\rangle+|11\rangle),\nonumber\\
|\epsilon_2\rangle\!&\!=\!&\!{1\over\sqrt{2}}(|01\rangle+|10\rangle),
\end{eqnarray}
with
\begin{equation}
\eta\;=\;{{\Delta+(\Delta^2+\chi^2)^{1/2}}\over\chi}.
\end{equation}
Initially preparing the system at
the (entangled) ground state $|\epsilon_1\rangle$, one can produce the
{\it maximally entangled} state $|\epsilon_2\rangle$ using a microwave-field pulse of
duration $\tau=\pi/\Omega_{12}$, where the Rabi frequency $\Omega_{12}$ is given by
$|t_{12}|/\hbar$ for a weak driving field.

At $f=0.5$, we have $\hbar\omega_{12}=0.00528E_J$, $\hbar\omega_{24}=0.03124E_J$, and
$|t_{24}/t_{12}|\approx 0.41$. When the microwave field is in resonance with the transition
$|\epsilon_1\rangle\rightarrow|\epsilon_2\rangle$ at $f=0.5$,
\begin{equation}
{\rho_{24}\over\rho_{12}}\approx 4.4\times 10^{-6}
\end{equation}
for 
\[
\tau=\pi/\Omega_{12}\sim 20\pi/\omega_{12}.
\]
Because the state leagage is very small, a fast two-bit operation can also be implemented 
using a microwave field.

\begin{figure}
\includegraphics[width=3.4in,
bbllx=26,bblly=165,bburx=578,bbury=760]{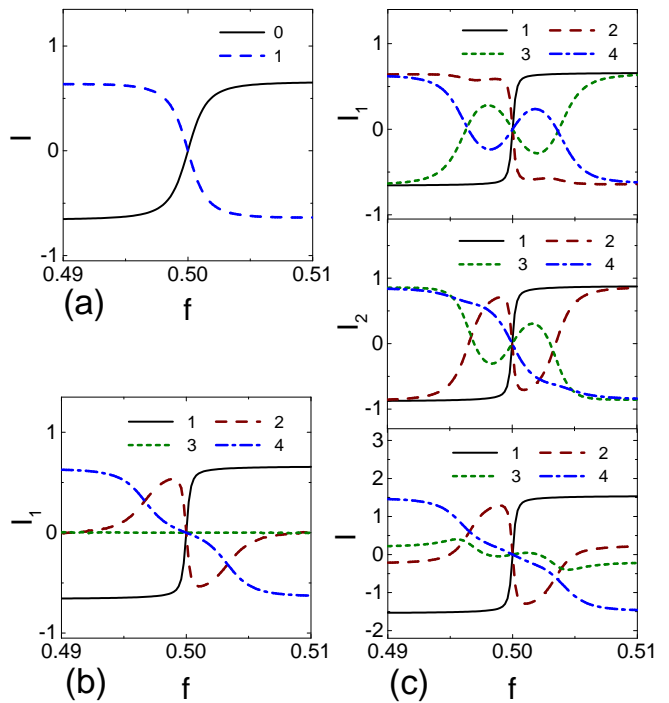}
\caption{(Color online) (a)~Supercurrents $I$ versus reduced flux $f$ at eigenstates $|0\rangle$
and $|1\rangle$ of a single flux qubit for $\beta_L=0.03$.
(b)~Supercurrents $I_1$ versus $f$ at eigenstates $|\epsilon_k\rangle$, $k=1$ to 4,
of two coupled flux qubits for the symmetric circuit with $\beta_{L1}=\beta_{L2}=0.03$.
(c)~Supercurrents $I_1$, $I_2$, and $I$ versus $f$ at eigenstates
$|\epsilon_k\rangle$, $k=1$ to 4, of two coupled flux qubits for 
an asymmetric circuit with $\beta_{L1}=0.03$
and $\beta_{L2}=0.04$. Here we choose $L_{b1}/L_c=0.1$ and $(L_{12}+L_{b2})/L_c=0.2$
for the coupled flux qubits. The supercurrents are in units of $I_0$ in (a) and $I_{01}$
in (b) and (c).
For the coupled two qubits, the total current $I=I_1+I_2$ is robust with respect to 
the asymmetry of $\beta_{L1}$ and $\beta_{L2}$.}
\label{fig7}
\end{figure}

\section{Supercurrents and quantum measurement}

The circulating supercurrents flowing through the inductance $L$
or $L_c$ are different for different eigenstates.
This property can be used for implementing a readout of the qubit states.
For a single flux qubit, around $f=0.5$,
the supercurrents $I$ at eigenstates $|0\rangle$ and $|1\rangle$
(i.e., $\langle 0|I|0\rangle$ and $\langle 1|I|1\rangle$) have equal magnitudes
but opposite directions [see Fig.~7(a)]. During quantum measurement,
one can switch on the flux transformer to couple the inductance $L$ with a dc-SQUID
magnetometer [cf. Fig.~1(a)] to distinguish the two eigenstates of the qubit
because at these two states the supercurrents $I$ through $L$ generate two different fluxes
in the SQUID loop of the magnetometer. In general, if the single flux qubit is
at the superposition state $c_1|1\rangle +c_0|0\rangle$, the measurement will show that
the qubit has probability $|c_i|^2$ at the eigenstate $|i\rangle$, where $i=0,1$.
For the two coupled flux qubits, the supercurrents through the common
inductance $L_c$ take different values at its four eigenstates.

Similar to the single flux qubit, a switchable flux transformer can be used to
couple $L_c$ and the SQUID loop of the magnetometer for reading out the
coupled-qubit states because the supercurrents $I$ at different eigenstates contribute
different fluxes in the SQUID loop of the magnetometer.
The supercurrents $I_1$ at the four eigenstates of the coupled qubits are shown
in Fig.~7(b) for two flux qubits having identical parameters. Since $I_1=I_2$ in this
case, the total supercurrent $I$ is $2I_1$. When the parameters of the two flux
qubits become {\it different}, the total supercurrents $I$ look {\it similar} to those
in Fig.~7(b), but $I_1$ and $I_2$ (which flow through the Josephson junctions of
the qubits) change {\it drastically} [cf. Fig.~7(c)].
Also, it can be seen that at the eigenstates of the system the circulating supercurrents
in both single and coupled flux qubits fall to zero at $f=0.5$.
To read out the qubit states, one can shift the system away from this point.

\section{Discussion and conclusion}

For the charge qubits coupled by $LC$-oscillator modes~\cite{MAK} or by
an inductance,\cite{YOU} the inductances proposed to be used are $\sim 3.6$~$\mu$H
or $\sim 30$~nH, respectively,
for a two-bit operation ten times slower than the
typical one-bit operation.
An inductance for coupling charge qubits similar to that in Ref.~\onlinecite{MAK},
particularly, has a value much larger than $L_c$ ($\approx 20$~pH) for
coupling flux qubits. It is difficult to fabricate in a small size without introducing
a strong coupling with the environment. Because two-bit operations are much slower than
one-bit ones in the inductively coupled {\it charge} qubits,
an efficient scheme is thus required to minimize the number of two-bit
(as opposed to one-bit) operations to obtain a conditional gate.\cite{YOU}
However, for inductively coupled {\it flux} qubits,
the above limitation in using two-bit operations for constructing a conditional gate is
removed since two-bit operations can be as fast as one-bit ones.
In this case, {\it any} schemes for constructing conditional gates become {\it efficient}
by minimizing the number of operations that are used (either one- or two-bit).
Note that the common inductance of $L_c\approx 20$~pH can produce a
strong interbit coupling.
As a result, two-bit operations as fast as one-bit ones can be achieved.
This common inductance is comparable to the loop inductance, $L\sim 10$~pH,
of the single flux qubit currently realized in experiments.

To couple several flux qubits, the inductances of all loops involved could be 
small, comparable to the loop inductance of a single flux qubit currently 
realized in experiments. This is the case we studied in the present paper, 
where two coupled flux qubits are considered. If a number of flux qubits 
are coupled, the inductances of some loops will become larger, but the 
common or shared inductance for producing the interbit coupling can still be chosen 
small (about 20 pH). If the circuits except for the line $A_1L_cB_1$ 
(corresponding to the common or shared inductance) could be screened from the 
environment (that is a big challenge for experimentalists for sure), 
the main noise would be due to the small common or shared inductance.  

In conclusion, we have proposed an experimentally realizable method for inductively
coupled flux qubits that can achieve two-bit operations performing as fast as one-bit ones.
We treat both single and coupled flux qubits with more realistic models including
the loop inductance.
Moreover, we show that the coupled flux qubits have novel flux-dependent behaviors
in the transitions between states. We find that the forbidden transitions in the coupled
two flux qubits become allowed (except for some specific values of the external flux)
when the parameters of the two qubits change from being initially equal to each other 
and then making these different.

\begin{acknowledgments}
We thank J.S. Tsai, T. Yamamoto, Yu. Pashkin, O. Astafiev, Y.X. Liu, L.F. Wei, T. Tilma, 
F. Wilhelm, Q. Niu, and S.C. Bernstein for discussions and comments.
This work was supported in part by the National Security Agency (NSA) and Advanced Research
and Development Activity (ARDA) under Air Force Office of Research (AFOSR)
contract number~F49620-02-1-0334, and by the National Science Foundation grant
No.~EIA-0130383.
J.Q.Y. was also supported by the National Natural Science Foundation
of China grant No.~10174075 and the Special Funds for Major State Basic Research of China  
grant No.~G2001CB3095.
\end{acknowledgments}

\appendix*\section{Series expansion of the interaction Hamiltonian}
The supercurrent $I_i$ flows through each of the three Josephson junctions in the
$i$th flux qubit, so $I_i$
can also be written as
\begin{eqnarray}
I_i\!&\!=\!&\!\alpha_iI_{0i}\sin\phi_3^{(i)} \nonumber\\
\!&\!=\!&\!-\alpha_iI_{0i}\sin\left\{2\pi(f'_i+I_jL_c/\Phi_0)+2\phi_{mi}\right\},
\end{eqnarray}
where $i,j=1,2$ and $i\ne j$. Taking advantage of this relation for $I_i$,
one can expand the interaction Hamiltonian (22) as
\begin{equation}
H_I=-\lambda L_cI_1I_2-\sum_{i=1}^2\alpha_iE_J^{(i)}\xi_i,
\end{equation}
where
\begin{equation}
\lambda=1+\sum_{i=1}^2\left[{1\over 3}\beta_{Li}^2\left(I_i\over I_{0i}\right)^2
+{2\over 15}\beta_{Li}^4\left(I_i\over I_{0i}\right)^4+\dots\right],
\end{equation}
and
\begin{eqnarray}
\xi_i\!&\!=\!&\!{1\over 2}\cos(2\pi f'_i+2\phi_{mi})\nonumber\\
&&\times\left[\beta_{Lj}^2\left(I_j\over I_{0j}\right)^2
+{5\over 12}\beta_{Lj}^4\left(I_j\over I_{0j}\right)^4+\dots\right],~~
\end{eqnarray}
with 
\begin{equation}
\beta_{Li}\equiv 2\pi I_{0i}L_c/\Phi_0<\pi/2.
\end{equation}
The term $-\lambda L_cI_1I_2$ in $H_I$ produces an interbit coupling between flux
qubits 1 and 2, while $\alpha_iE_J^{(i)}\xi_i$ slightly modifies the energy
levels of the $i$th flux qubit.


\end{document}